# Dielectric properties of $Li_2O$-$3B_2O_3$ glasses

Rahul Vaish and K. B. R. Varma*

Materials Research Centre, Indian Institute of Science, Bangalore-560 012, India.

*Corresponding Author; E-Mail: kbrvarma@mrc.iisc.ernet.in;

FAX: 91-80-23600683; Tel. No: 91-80-22932914

**Abstract:**

The frequency and temperature dependence of the dielectric constant and the electrical conductivity of the transparent glasses in the composition $Li_2O$-$3B_2O_3$ (LBO) were investigated in the 100 Hz- 10 MHz frequency range. The dielectric constant and the loss in the low frequency regime were electrode material dependent. Dielectric and electrical relaxations were respectively analyzed using the Cole-Cole and electric modulus formalisms. The dielectric relaxation mechanism was discussed in the framework of electrode and charge carrier (hopping of the ions) related polarization using generalized Cole-Cole expression. The frequency dependent electrical conductivity was rationalized using Jonscher's power law. The activation energy associated with the dc conductivity was 0.80 ± 0.02 eV, which was ascribed to the motion of $Li^+$ ions in the glass matrix. The activation energy associated with dielectric relaxation was almost equal to that of the dc conductivity, indicating that the same species took part in both the processes. Temperature dependent behavior of the frequency exponent ($n$) suggested that the correlated barrier hopping model was the most apposite to rationalize the electrical transport phenomenon in $Li_2O$-$3B_2O_3$ glasses. These glasses on heating at 933 K/10h resulted in the known non-linear optical phase $LiB_3O_5$.

## 1. Introduction:

Noncentrosymmetric borate-based compounds have been becoming increasingly important, because of their symmetry dependent properties such as piezoelectric, pyroelectric, ferroelectric and non-linear optical. Various borate-based single crystals, including $BaB_2O_4$ [1], $BiB_3O_6$ [2], $SrB_4O_7$ [3], $CsLiB_6O_{10}$ [4], $LiNaB_4O_7$ [5] have been investigated and reported to be promising from their physical properties view point. These compounds have attained importance particularly for their use in non-linear optical devices.

Amongst various borate-based materials, $LiB_3O_5$ single crystals have been reported to be promising from their non-linear optical, pyroelectric and piezoelectric properties viewpoint [6-8]. It has high non-linear optic coefficient, large optical damage threshold, wide transmission window, low cost associated with fairly good chemical and mechanical stability. $LiB_3O_5$ belongs to the orthorhombic crystal class associated with $Pna2_1$ space group. Because of the combination of the promising physical properties indicated above, $LiB_3O_5$ as a functional material in the $Li_2O$-$B_2O_3$ binary system has attracted the attention of several researchers around the globe. The same binary system also yields $Li_2B_4O_7$ compound which is technologically important owing to its interesting surface acoustic wave (SAW), piezoelectric and pyroelectric properties [9,10].

Though, the optical properties of $LiB_3O_5$ have been studied in detail [11, 12], the literature on the electric properties is limited [6]. Since $LiB_3O_5$ is polar, it deserves much attention from its electrical transport properties point of view as these properties have direct influence on its pyroelectric and piezoelectric characteristics. Glass-ceramic route of fabricating transparent materials at finer scale which eventually on heat treatment yields the desired crystalline phase is an industrially important method. Therefore, to begin with, glasses in the composition $Li_2O$-$3B_2O_3$ (which on heating at appropriate temperatures yielded crystalline $LiB_3O_5$ phase) have been investigated for their dielectric and electrical conductivity properties over the range of temperatures and frequencies that are normally of interest in the applications of these materials. The experimental data have been modeled using Cole-Cole relation [13], Jonscher's power law [14] and electric modulus formalism. The details pertaining to these studies are reported in the following sections.

**2. Experimental:**

Transparent glasses in the composition $Li_2O$-$3B_2O_3$ (LBO) were fabricated via the conventional melt-quenching technique. For this, $Li_2CO_3$ and $H_3BO_3$ were mixed and melted in a platinum crucible at 1173 K for 30 min. The batch weight was 20 gm. Melts were quenched by pouring on a steel plate and pressed with another plate to obtain 1-2 mm thick glass plates. These glasses were annealed at 673 K for 12 h. The amorphous nature of the as-quenched samples was confirmed by X-ray powder diffraction (XRD, Philips PW1050/37) using Cu Kα radiation. The glassy characteristics were established by differential scanning calorimetry (DSC, Model: Diamond DSC, Perkin Elmer) studies.

The capacitance and dielectric loss ($D$) measurements on the as-quenched (annealed) polished glass plates of 1 and 3 mm in thickness using various electrode materials (Ag-paint, sputtered Au and thermally evaporated Al) were done using impedance gain phase analyzer (HP 4194 A) in the 100 Hz-10 MHz frequency range with a signal strength of 0.5 $V_{rms}$ at various temperatures (300–525 K). Thin silver leads were bonded to the sample using silver epoxy. Based on these data the dielectric constants were evaluated by taking the dimensions and electrode geometry of the sample into account.

**3. Results and Discussion:**

The DSC trace that was obtained in the 600- 950 K temperature range for the as-quenched glass plates is shown in Fig. 1. It exhibits the glass transition (endotherm, 770 K) and exotherms in the 835 -910 K temperature range associated with the crystallization. The XRD pattern obtained for the as-quenched sample ($Li_2O$-$3B_2O_3$ in molar ratio) that is shown in Fig.2 (a) confirms its amorphous state. In order to ascertain the crystalline phases that are evolving in the above temperature range, the as-quenched samples were heated to 823 K/4 h, 873 K/6 h and 933 K/10 h. The X-ray powder diffraction pattern obtained for the 823 K/4 h heat-treated samples is depicted in Fig. 2 (b). The Bragg peaks that are encountered in this pattern could be indexed to $Li_2B_4O_7$ (major phase) and $Li_2B_8O_{13}$ (minor phase) phases. The XRD pattern [Fig. 2 (c)] that was obtained for the 873 K/6 h heat-treated sample also revealed the presence of two major phases corresponding to $Li_2B_4O_7$ and $Li_2B_8O_{13}$. In order to ascertain the thermal stability of these phases ($Li_2B_4O_7$ and $Li_2B_8O_{13}$), the as-quenched glasses were heat-treated at 933 K/10 h (beyond the intense exotherm in Fig. 1). The XRD pattern that was obtained at room temperature for this sample is shown in Fig. 2 (d). It is interesting to note that all the Bragg peaks in this pattern could be assigned to $LiB_3O_5$ phase (a=8.446(2) Å, b=7.380(2) Å, and c= 5.147(2) Å). The as-quenched glasses are likely to have diborate and tetraborate structural units which favor crystallization of $Li_2B_4O_7$ and $Li_2B_8O_{13}$ at the initial stages of crystallization [15]. Based on the XRD studies, the scheme of the crystallization of $Li_2O$-$3B_2O_3$ glasses may be illustrated as follows;

$$Li_2O\text{-}3B_2O_3 \xrightarrow{873K/6h} 0.5Li_2B_4O_7 + 0.5Li_2B_8O_{13} \xrightarrow{933K/10h} 2LiB_3O_5 \quad (1)$$

The variation of the dielectric constant ($\varepsilon_r^{'}$) with frequency (100 Hz – 10 MHz) of measurement for 1 mm thick LBO glass-plates (with silver paint electrodes) at different temperatures is shown in Fig. 3. At all the temperatures under investigation, $\varepsilon_r^{'}$ decreases with increase in frequency. The decrease is significant, especially at low frequencies, which may be associated with the mobile ion polarization combined with electrode polarization. The low-frequency dispersion of $\varepsilon_r^{'}$ gradually increases with increase in temperature due to an increase in the electrode polarization as well as the thermal activation associated with $Li^+$ ions in the LBO glasses. The electrode polarization is significant at high temperatures (423 K-523 K) and masks the bulk response of the glasses in the low frequency regime. When the temperature rises, the dielectric dispersion shifts towards higher frequencies.

To begin with, an attempt was made to rationalize the dielectric relaxation in LBO glasses by using the Cole-Cole equation [13]:

$$\varepsilon_r^* = \varepsilon_\infty + \frac{\varepsilon_s - \varepsilon_\infty}{1 + (i\omega\tau)^{1-\alpha}} \quad (2)$$

where $\varepsilon_s$ is the static dielectric constant, $\varepsilon_\infty$ is a high frequency value of the dielectric constant, $\omega$ (=2$\pi f$) is the angular frequency, $\tau$ is the dielectric relaxation time and $\alpha$ is a measure of distribution of relaxation times with values ranging from 0 to 1. For an ideal Debye relaxation, $\alpha$ = 0 and $\alpha$ > 0 indicates that the relaxation has a distribution of relaxation times. After solving Eq. 2 for the dielectric constant, one obtains

$$\varepsilon_r' = \varepsilon_\infty + \frac{(\varepsilon_s - \varepsilon_\infty)[1 + (\omega\tau)^{1-\alpha}\sin(\alpha\pi/2)]}{1 + 2(\omega\tau)^{1-\alpha}\sin(\alpha\pi/2) + (\omega\tau)^{2-2\alpha}} \tag{3}$$

The experimental data on the variation of $\varepsilon_r'$ with frequency could not be fitted perfectly using Eq. 3 in the entire frequency range since the Cole-Cole equation predicts nearly constant $\varepsilon_r'$ in the low frequency regime, which is not true in the present case. This is due to the fact that the electrode/space charge polarization is dominant at low frequencies as depicted in Fig. 3. The above observations necessitate the inclusion of the electrical conductivity term in the Cole-Cole equation to rationalize the $\varepsilon_r'$ versus frequency behavior of LBO glasses in the whole frequency range. After adding the term that reflects the electrode/space charge polarization in the Eq. 3, one arrives at [16]

$$\varepsilon_r' = \varepsilon_\infty + \frac{(\varepsilon_s - \varepsilon_\infty)[1 + (\omega\tau)^{1-\alpha}\sin(\alpha\pi/2)]}{1 + 2(\omega\tau)^{1-\alpha}\sin(\alpha\pi/2) + (\omega\tau)^{2-2\alpha}} + \frac{\sigma_2}{\varepsilon_o\omega^s} \tag{4}$$

where $s$ (0, 1) is a constant and $\sigma_2$ is the conductivity, a contribution from the space charges. Solid lines in Fig. 3 are the fitted curves (Goodness of fit ($R^2$) >0.999) of the experimental results (100 Hz-10 MHz) according to Eq. 4. The parameters that are obtained from the best fit at various temperatures are presented in table I. In order to further elucidate the dielectric relaxation in LBO glasses, it is important to estimate the activation energy associated with the relaxation process. The activation energy involved in the relaxation process of ions could be obtained from the temperature dependent relaxation time (table I) as

$$\tau = \tau_o \exp\left(E/kT\right) \tag{5}$$

where $E$ is the activation energy associated with the relaxation process, $\tau_o$ is the pre-exponential factor, $k$ is the Boltzmann constant, and $T$ is the absolute temperature. Fig. 4 depicts the plot of ln (τ) versus 1000/*T* along with linear fit (solid line) to the above equation (Eq. 5). The value that is obtained for *E* is 0.77 ± 0.03eV, which is ascribed to the motion of $Li^+$ ions [17] in the glass matrix.

The variation of the dielectric loss (*D*) with the frequency at various temperatures is shown in Fig. 5. The loss decreases with increase in frequency at different temperatures (313 K-523 K). However, it increases with increase in temperature, which is attributed to the increase in electrical conductivity of the glasses. A relaxation peak at 150 Hz was encountered when the measurements were done at 523 K. In order to understand the effect of electrode materials used on the dielectric relaxation of LBO glasses in the low frequency regime, different electrode materials (silver paint, sputtered gold and thermally evaporated aluminium) were used. Fig. 6 (a & b) shows the dielectric constant and the loss behavior at 523 K for various electrode materials. Significant difference in the dielectric constants for different electrode materials was observed at low frequencies [Fig. 6 (a)]. However, all the plots merge in the high frequency regime (above 10 kHz). This electrode independent behavior at high frequencies (10 kHz-10 MHz) is attributed to the intrinsic dielectric response of the glasses. Interestingly, clear relaxation peaks were observed in the frequency dependent dielectric loss plots [Fig. 6 (b)]. The frequency associated with the dielectric relaxation was found to vary with the electrode materials used suggesting that the above relaxation is ascribed to electrode polarization. All the plots overlap in the high frequency region akin to that of the dielectric constant plots [Fig. 6 (a)]. Inorder to probe further into these results the dielectric measurements were performed on the samples of two different thicknesses (1 mm and 3 mm) using silver paint electrodes at 523 K (Fig. 7). The dielectric dispersion is found to be dependent on the thickness of the sample indicating space charge polarization at sample/electrode interfaces contributing to the observed dielectric dispersion.

Electric modulus formalism was also invoked to rationalize the dielectric response of the present glasses. The use of electric modulus approach helps in understanding the bulk response of moderately conducting samples. This would facilitate to circumvent the problems caused by electrical conduction which might mask the dielectric relaxation processes. The complex electric modulus ($M^*$) is defined in terms of the complex dielectric constant ($\varepsilon$*) and is represented as [18]:

$$M^* = (\varepsilon^*)^{-1} \quad (6)$$

$$M^* = M' + iM'' = \frac{\varepsilon_r'}{(\varepsilon_r')^2 + (\varepsilon_r'')^2} + i\frac{\varepsilon_r''}{(\varepsilon_r')^2 + (\varepsilon_r'')^2} \quad (7)$$

where $M'$, $M''$ and , $\varepsilon_r'$, $\varepsilon_r''$ are the real and imaginary parts of the electric modulus and dielectric constants, respectively. The real and imaginary parts of the modulus at different temperatures are calculated using Eq. 7 for the LBO glasses and depicted in Figs. 8 (a & b), respectively. One would notice from Fig. 8 (a) that at low frequencies, $M'$ approaches zero at all the temperatures under study suggesting the suppression of the electrode polarization. $M'$ reaches a maximum value corresponding to $M_\infty = (\varepsilon_\infty)^{-1}$ due to the relaxation process. It is also observed that the value of $M_\infty$ decreases with the increase in temperature. The imaginary part of the electric modulus (Fig. 8 (b)) is indicative of the energy loss under electric field. The $M''$ peak shifts to higher frequencies with increasing temperature. This evidently suggests the involvement of temperature dependent relaxation processes in the present glasses. The frequency region below the $M''$ peak indicates the range in which $Li^+$ ions drift to long distances. In the frequency range which is above the peak, the ions are spatially confined to potential wells and free to move within the wells. The frequency range where the peak occurs is suggestive of the transition from long-range to short-range mobility. The electric modulus ($M^*$) could be expressed as the Fourier transform of a relaxation function ϕ(t):

$$M^* = M_\infty \left[ 1 - \int_0^\infty \exp(-\omega t) \left( -\frac{d\phi}{dt} \right) dt \right] \tag{8}$$

where the function ϕ(t) is the time evolution of the electric field within the materials and is usually taken as the Kohlrausch-Williams-Watts (KWW) function [19,20]:

$$\phi(t) = \exp\left[ -\left( t/\tau_m \right)^\beta \right] \tag{9}$$

where $\tau_m$ is the conductivity relaxation time and the exponent $\beta$ (0 1] indicates the deviation from Debye-type relaxation. The value of $\beta$ could be determined by fitting the experimental data in the above equations. But it is desirable to reduce the number of adjustable parameters while fitting the experimental data. Keeping this point in view, the electric modulus behavior of the present glass system is rationalized by invoking modified KWW function suggested by Bergman. The imaginary part of the electric modulus ($M^"$) is defined as [21]:

$$M" = \frac{M^"_{Max}}{(1-\beta) + \frac{\beta}{1+\beta}\left[ \beta(\omega_{Max}/\omega) + (\omega/\omega_{Max})^\beta \right]} \tag{10}$$

where $M^"_{Max}$ is the peak value of the $M^"$ and $\omega_{Max}$ is the corresponding frequency. The above equation (Eq. 10) could effectively be described for $\beta \geq 0.4$. Theoretical fit of Eq. 10 to the experimental data is shown in Fig. 8 (b) as the solid lines. It is seen that the experimental data are well fitted to this model except in the high frequency regime. From the fitting of $M^{''}$ versus frequency plots, the value of $\beta$ was determined and found to be temperature dependent. The plot of $\beta$ versus temperature is depicted in Fig. 9. $\beta$ increases gradually with the increase in temperature indicating that as the temperature increases the glass network loosens and the interactions between $Li^+$ ions and surrounding matrix decreases.

The relaxation frequency associated with this process was determined from the plot of $M^{''}$ versus frequency. The activation energy involved in the relaxation process of ions could be obtained from the temperature dependent frequency associated with the peak of $M^{''}$ as:

$$f_m = f_o \exp\left(-\frac{E_R}{kT}\right) \tag{11}$$

where $E_R$ is the activation energy associated with the relaxation process, $f_o$ is the pre-exponential factor, $k$ is the Boltzmann constant and $T$ is the absolute temperature. Fig. 10 shows a plot between ln ($f_m$) and 1000/$T$ along with the theoretical fit (solid line) to the above equation (Eq. 11). The value that is obtained for $E_R$ is 0.80 ± 0.02eV, which is ascribed to the motion of $Li^+$ ions and is consistent with the one reported in the literature [17].

In order to elucidate the electrical transport mechanism in LBO glasses, DC conductivity at different temperatures ($\sigma_{DC}(T)$), was calculated from the electric modulus data. The DC conductivity could be obtained according to the expression [22]:

$$\sigma_{DC}(T) = \frac{\varepsilon_o}{M_\infty(T) * \tau_m(T)}\left[\frac{\beta(T)}{\Gamma\left(1/\beta(T)\right)}\right] \tag{12}$$

where $\varepsilon_o$ is the free space dielectric constant, $M_\infty(T)$ is the reciprocal of high frequency dielectric constant and $\tau_m(T)$ ($=1/2\pi f_m$) is the temperature dependent relaxation time. Fig. 11 shows the DC conductivity data obtained from the above expression (Eq. 12) at various temperatures. The activation energy for the DC conductivity was calculated from the plot of ln ($\sigma_{DC}$) versus 1000/$T$ for LBO glasses, which is shown in Fig. 11. The plot is found to be linear and fitted using the following Arrhenius equation,

$$\sigma_{DC}(T) = B\exp\left(-E_{DC}/kT\right) \tag{13}$$

where $B$ is the pre-exponential factor, $E_{DC}$ is the activation energy for the DC conduction. The activation energy calculated from the slope of the fitted line is found to be 0.79 ± 0.03eV. This value of activation energy is higher than that of the value associated with dc conduction in $Li_2O$-$2B_2O_3$ glasses [23]. This is due to fact that in the alkali borate systems, the local structure of boron could be tailored by varying the alkali oxide content. At higher alkali content, more number of non-bridging oxygens (NBOs) are formed which yield a open structure of the borate network [24]. $Li_2O$-$2B_2O_3$ glasses have higher molar content of $Li_2O$ than that of the $Li_2O$-$3B_2O_3$ glasses which consequences the change in the coordination of boron associated with the formation of NBOs. The environment around $Li^+$ is changed due to variation in the NBOs. Such structural changes can have important influence on the mobility of $Li^+$ ions. The $Li^+$ mobility increases in the presence of NBOs. This suggests that the glasses in the composition of $Li_2O$-$2B_2O_3$ would have higher conductivity than that of the $Li_2O$-$3B_2O_3$ glasses.

AC conductivity at different frequencies and temperatures, was determined by using the dielectric data using the following formula:

$$\sigma_{AC} = \omega\varepsilon_o D\varepsilon_r' \tag{14}$$

where $\sigma_{AC}$ is the AC conductivity at a frequency $\omega$ (=2π*f*). The frequency dependence of the AC conductivity at different temperatures is shown in Fig. 12. At low frequency, the conductivity shows a flat response which corresponds to the dc part of the conductivity. At higher frequencies, the conductivity shows a dispersion. It is clear from the figure that the flat region increases with the increase in temperature. The phenomenon of the conductivity dispersion in solids is generally analyzed using Jonscher's law

$$\sigma_{AC} = \sigma_{DC} + A\omega^n \tag{15}$$

where $\sigma_{DC}$ is the dc conductivity, *A* is the temperature dependent constant and *n* is the power law exponent which generally varies between 0 and 1. The exponent *n* represents the degree of interaction between the mobile ions. The present glasses are found to obey the above mentioned universal power law at all the temperatures and frequencies under study. The theoretically fitted lines of Eq. 15 to the experimental data are shown in Fig. 12 (solid lines). The conductivity obtained for the present glasses at 500 Hz and 373 K is 1.5 X $10^{-7}$ $\Omega^{-1}.m^{-1}$ which is in the same order of magnitude for $LiB_3O_5$ single crystals along a and b-axes (9.5 X $10^{-8}$ $\Omega^{-1}.m^{-1}$) [6]. However, slightly higher value of conductivity associated with $Li_2O$-$3B_2O_3$ glass is attributed to the easy migration of $Li^+$ ions through the diborate and tetraborate structural units.

The variation of exponent *n* as a function of temperature is depicted in Fig. 9. It is known that the conductivity mechanism in any material could be understood from the temperature dependent behavior of *n*. To ascertain the electrical conduction mechanism in the materials, various models have been proposed [25]. These models include quantum mechanical tunneling model (QMT), the overlapping large-polaron tunneling model (OLPT) and the correlated barrier hopping model (CBH). According to the QMT model, the value of exponent $n$ is found to be 0.8 and increases slightly with increase in the temperature whereas the OLPT model predicts the frequency and temperature dependence of $n$. In the CBH model, the temperature dependent behavior of $n$ is proposed. This model states that the charge transport between localized states due to hopping over the potential barriers and predicts a decrease in the value of $n$ with the increase in temperature, which is consistent with the behavior of $n$ for the glasses understudy (Fig. 9). This suggests that the conductivity behavior of LBO glasses can be explained using correlated barrier hopping model.

The present glasses do not seem to follow the Ngai's relation ($\beta$ =1-$n$) [26] as the plots of imaginary part of electric modulus are not fitted exactly in the high frequency regime which influences the value of $\beta$ [Fig. 8 (b)]. Since the values for $\beta$ and $n$ are estimated in different frequency regions (as they could not be fitted well in the same frequency region), it is inconsistent with the Ngai's relation. Although the qualitative changes in the values of $\beta$ and $n$ are in conformity with the fact that both parameters represent the interaction between the ions [27].

The temperature dependence of the AC conductivity at different frequencies is shown in Fig. 13. At high temperatures and low frequencies the curves tend to merge with each other with a constant slope. This frequency independent behavior is attributed to the contribution from the DC conduction. The solid line that is shown in Fig. 13 is the linear fit. The slope of which gives the activation energy which is about 0.82 ± 0.03eV attributed to the $Li^+$ ion transport. It is worth noting that the activation energies for relaxation process and DC conduction are in close agreement. It suggests that similar energy barriers are involved in both the relaxation and conduction processes.

**4. Conclusions:**

The frequency and temperature dependence of dielectric properties of $Li_2O$-$3B_2O_3$ glasses were investigated in the frequency range of 100 Hz - 10 MHz. The dielectric relaxation peak was observed in the frequency dependent dielectric loss plots whose magnitude had electrode materials dependence. The dielectric relaxation behavior of these glasses was rationalized using Cole-Cole equation and the electrical transport properties were investigated and found to be obeying Jonscher's universal law. The activation energy associated with the dielectric relaxation determined from the dielectric and electric modulus spectra was found to be 0.78 ± 0.04 eV, close to that the activation energy for DC conductivity (0.80 ± 0.02 eV).

Table I: Fitted parameters from the Cole-Cole equation for $Li_2O$-$3B_2O_3$ glasses.

| $T$ (K) | $\varepsilon_s$ | $\varepsilon_\infty$ | $\tau$ (μs) | α | $\sigma_2$ $(\Omega..m)^{-1}$ | $s$ |
|---|---|---|---|---|---|---|
| 423 | 19 | 8.5 | 80 | 0.428 | 7.8 E-10 | 0.39 |
| 448 | 19.8 | 8.8 | 20 | 0.401 | 2.31E-9 | 0.53 |
| 473 | 20.7 | 9 | 5.6 | 0.34 | 2.17E-7 | 0.98 |
| 498 | 21.5 | 9.4 | 2.63 | 0.336 | 5.85E-7 | 0.99 |
| 523 | 22 | 9.85 | 1.46 | 0.30 | 5.25E-6 | 0.99 |

**Figure captions:**

Fig. 1: DSC trace for as-quenched $Li_2O$-$3B_2O_3$ glass plates.

Fig. 2: X-ray powder diffraction patterns for the (a) as-quenched, (b) 823 K/4 h, (c) 873 K/6 h and (d) 933 K/10 h heat-treated $Li_2O$-$3B_2O_3$ glasses.

Fig. 3: Frequency dependent dielectric constant plots at various temperatures and solid lines are the fitted curves using Eq. 3 in the text.

Fig. 4: ln ($\tau$) versus $1000/T$ plot for $Li_2O$-$3B_2O_3$ glasses.

Fig. 5: Dielectric loss versus frequency plots at various temperatures.

Fig. 6: Frequency dependent behavior of (a) Dielectric constant and (b) dielectric loss using various electrode materials at 523 K.

Fig. 7: Variation in dielectric constant with frequency for the samples of two different thicknesses (1 mm and 3 mm).

Fig. 8: (a) Real and (b) imaginary parts of the electric modulus as a function of frequency at various temperatures. The solid lines are the theoretical fits.

Fig. 9: $n$ & $\beta$ versus $T$ for $Li_2O$-$3B_2O_3$ glasses.

Fig. 10: Arrhenius plot for electrical relaxation.

Fig. 11: Arrhenius plot for DC conductivity.

Fig. 12: Variation of AC conductivity as a function of frequency at different temperatures and solid lines are the fitted curves.

Fig. 13: Temperature dependence of AC conductivity at different frequencies and solid line is the linear fit.

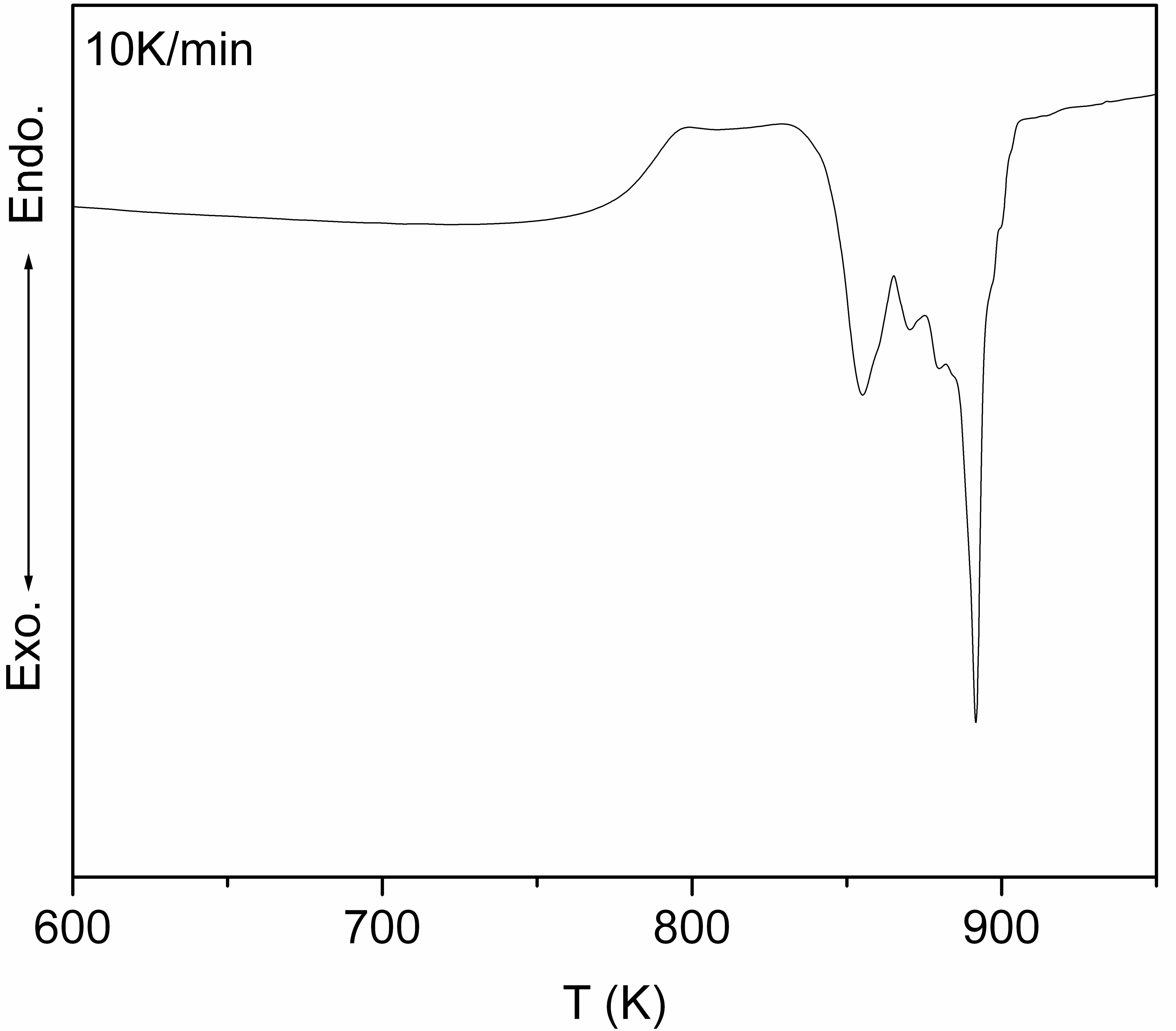

10K/min
Endo.
Exo.
600
700
800
900
T (K)

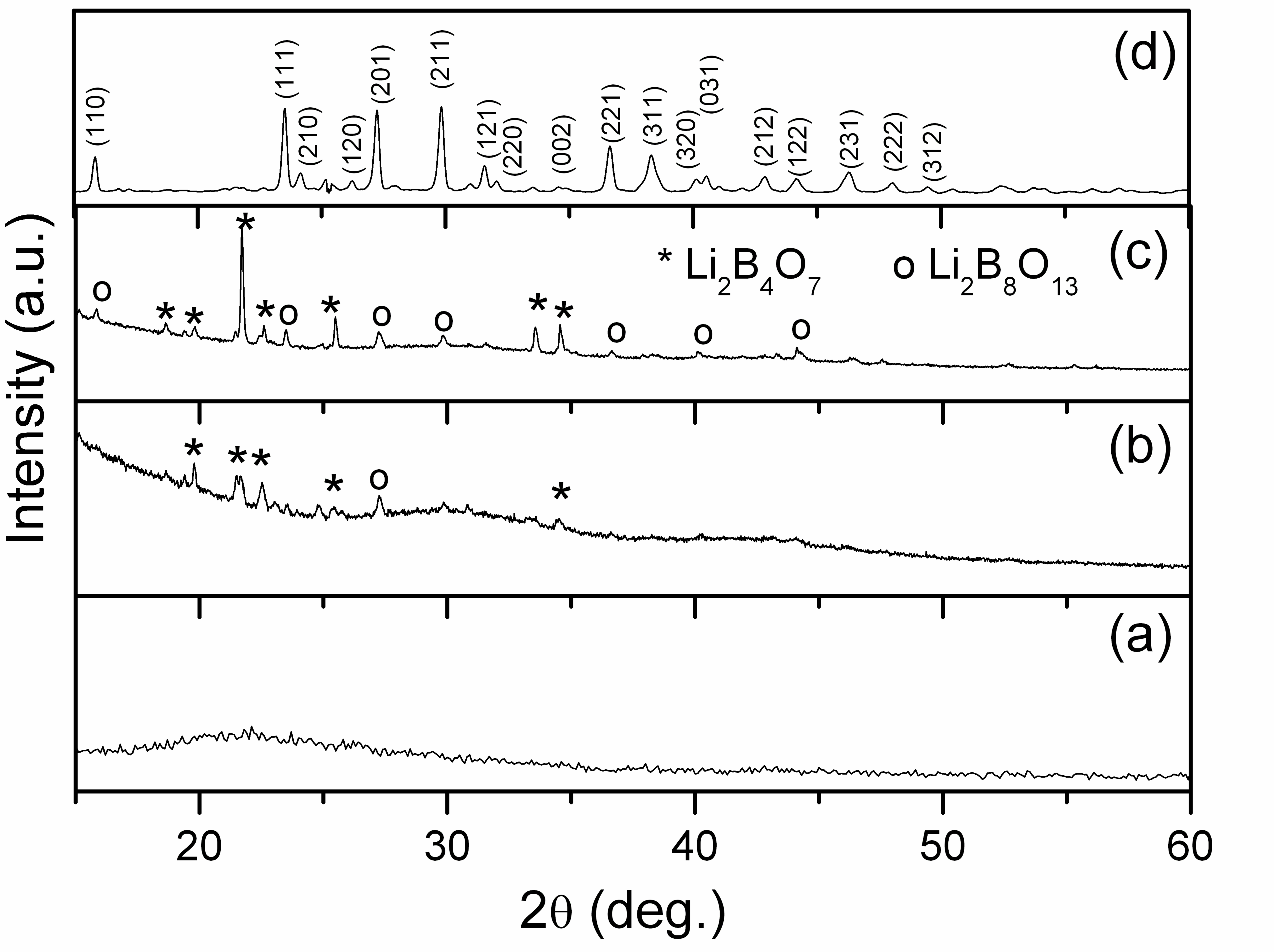

(d)
(110)
(111)
(210)
(120)
(201)
(211)
(121)
(220)
(002)
(221)
(311)
(320)
(031)
(212)
(122)
(231)
(222)
(312)
(c)
* Li2B4O7
o Li2B8O13
(b)
(a)
Intensity (a.u.)
20
30
40
50
60
2θ (deg.)

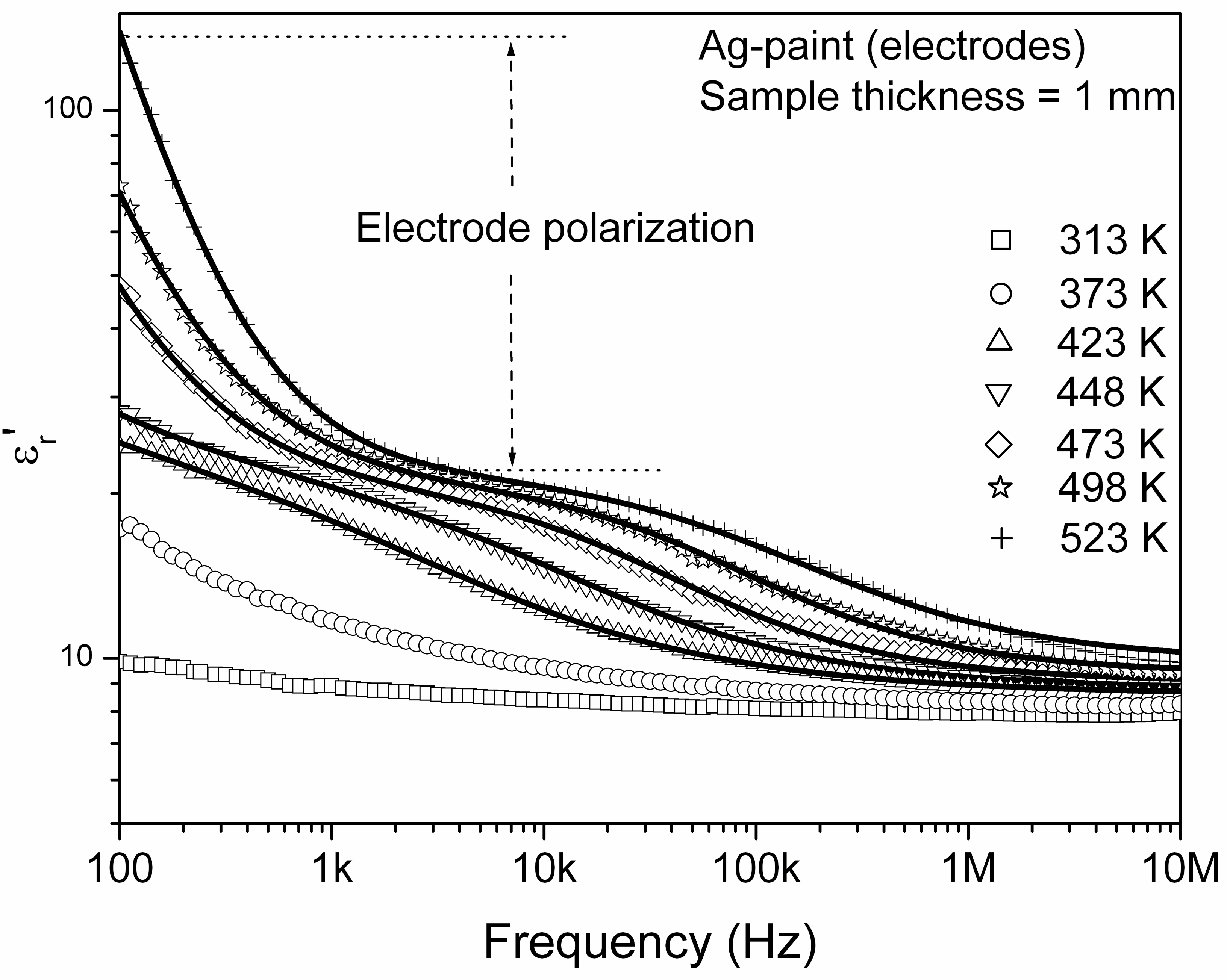

Ag-paint (electrodes)
Sample thickness = 1 mm
Electrode polarization
313 K
373 K
423 K
448 K
473 K
498 K
523 K
ε_r'
100
10
100
1k
10k
100k
1M
10M
Frequency (Hz)

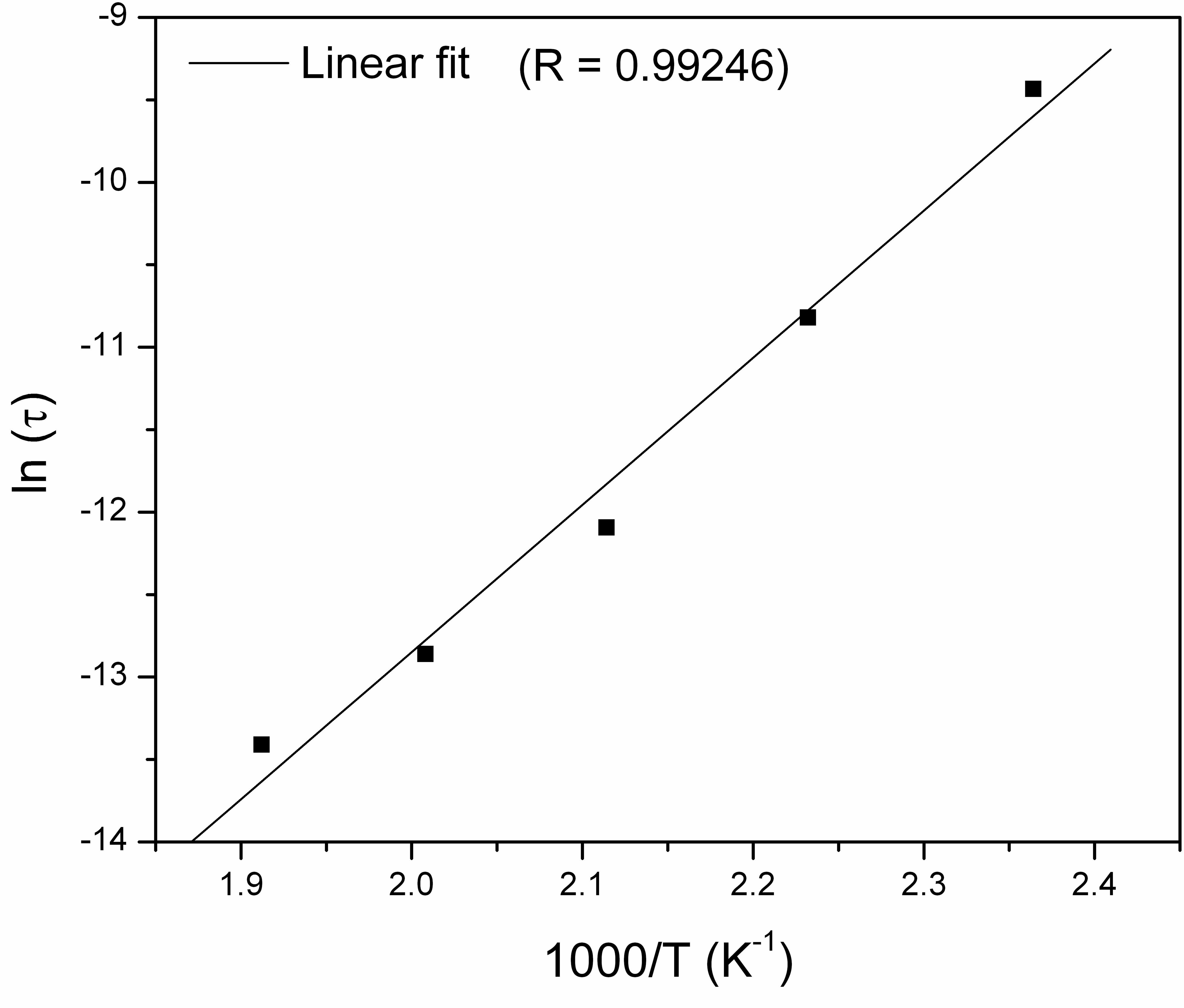

Linear fit (R = 0.99246)
ln (τ)
1000/T (K^-1)
-9
-10
-11
-12
-13
-14
1.9
2.0
2.1
2.2
2.3
2.4

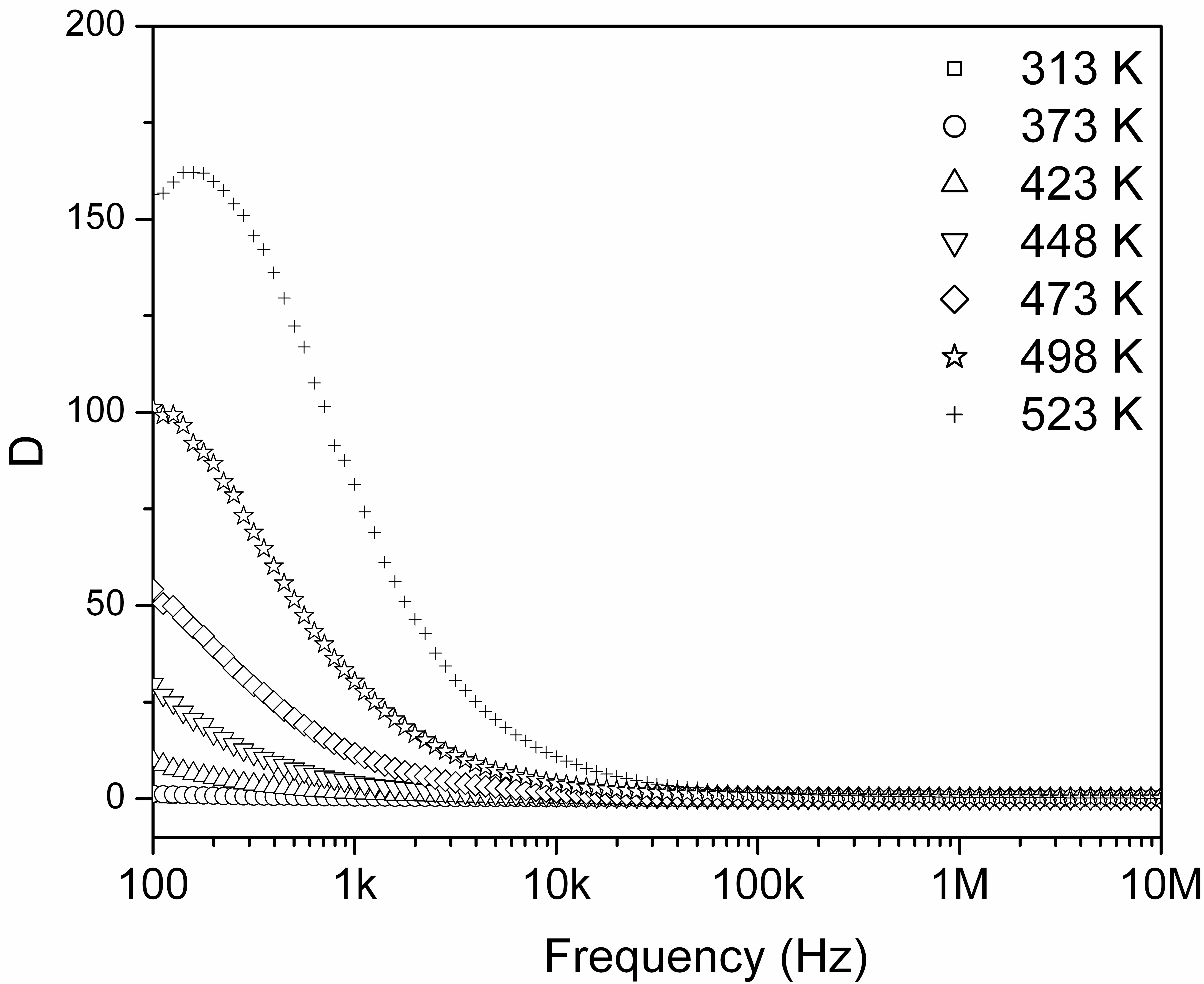

313 K
373 K
423 K
448 K
473 K
498 K
523 K
D
Frequency (Hz)
200
150
100
50
0
100
1k
10k
100k
1M
10M

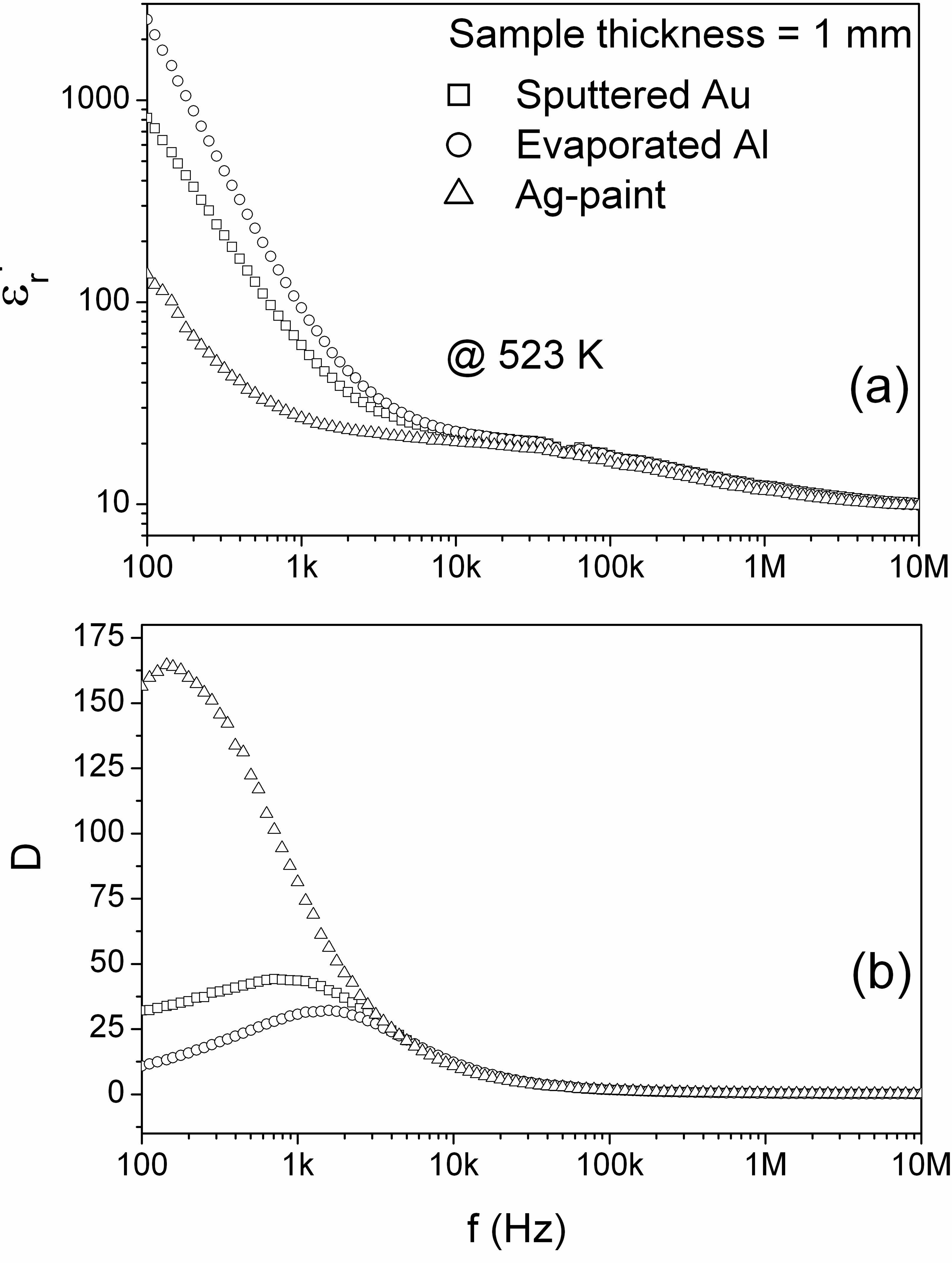
Sample thickness = 1 mm
Sputtered Au
Evaporated Al
Ag-paint
@ 523 K
(a)
$\varepsilon_r'$
1000
100
10
100
1k
10k
100k
1M
10M
(b)
D
175
150
125
100
75
50
25
0
100
1k
10k
100k
1M
10M
f (Hz)

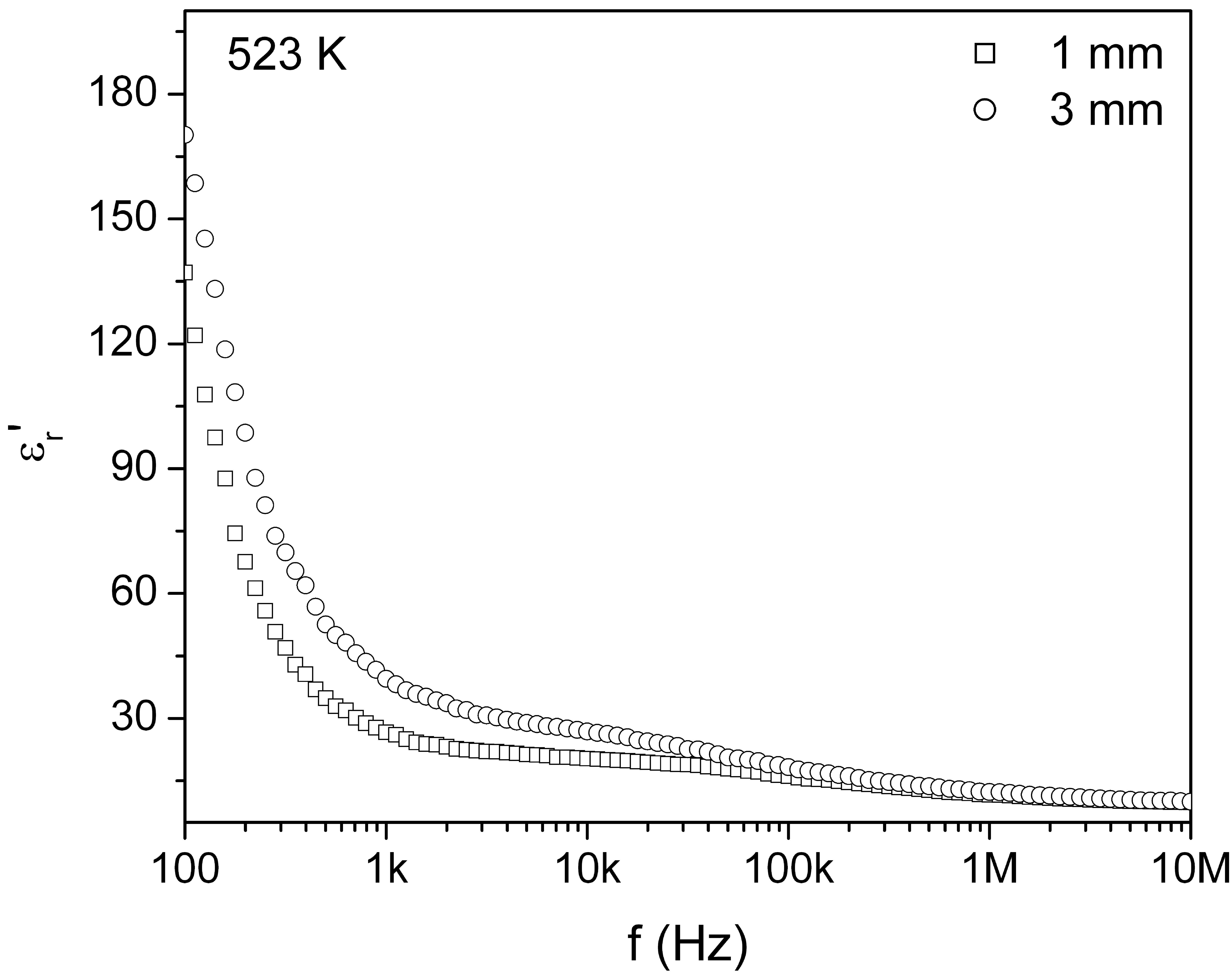

523 K
1 mm
3 mm
180
150
120
90
60
30
ε'r
100
1k
10k
100k
1M
10M
f (Hz)

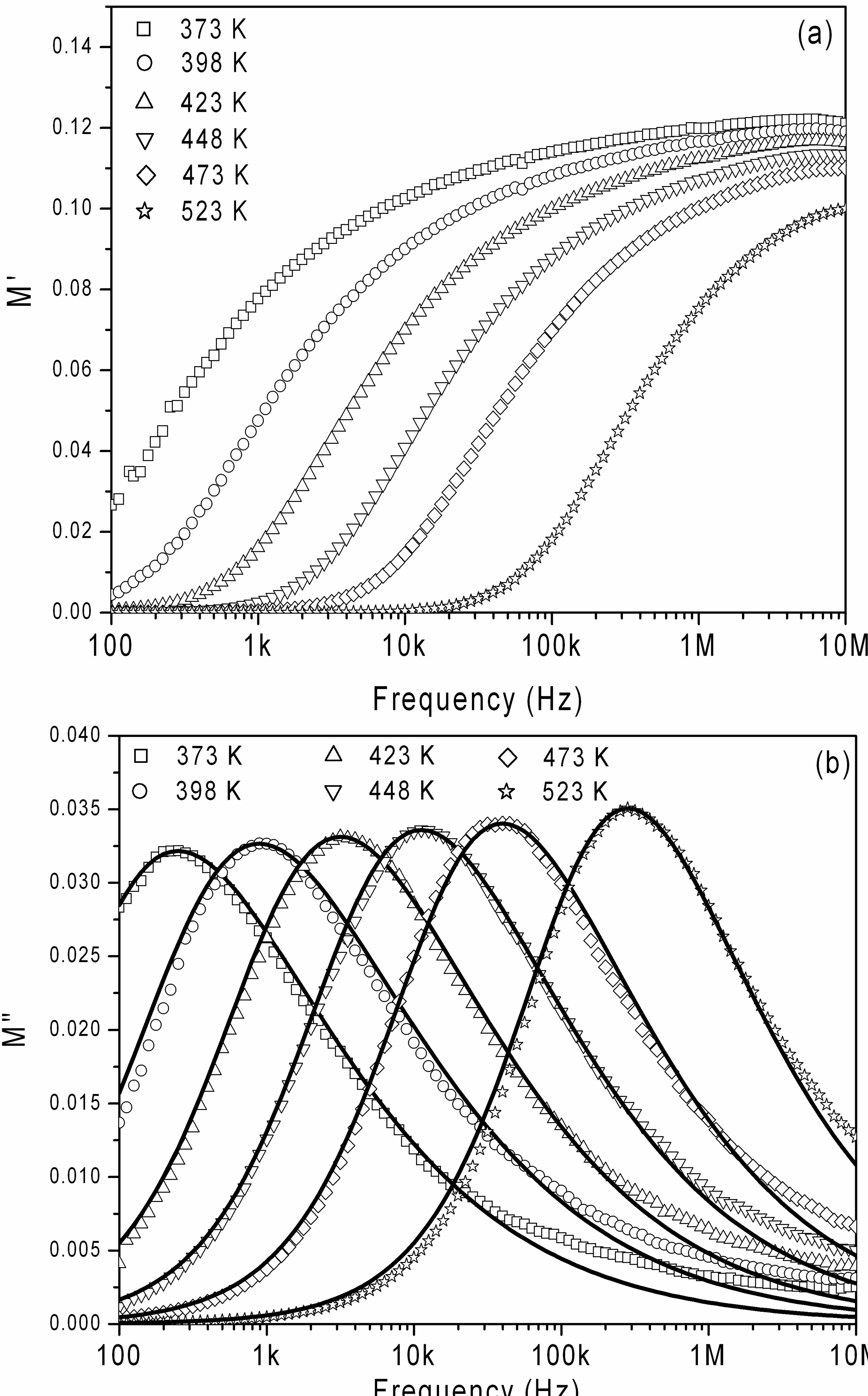

(a)
373 K
398 K
423 K
448 K
473 K
523 K
M'
0.00
0.02
0.04
0.06
0.08
0.10
0.12
0.14
100
1k
10k
100k
1M
10M
Frequency (Hz)
(b)
373 K
398 K
423 K
448 K
473 K
523 K
M''
0.000
0.005
0.010
0.015
0.020
0.025
0.030
0.035
0.040
100
1k
10k
100k
1M
10M
Frequency (Hz)

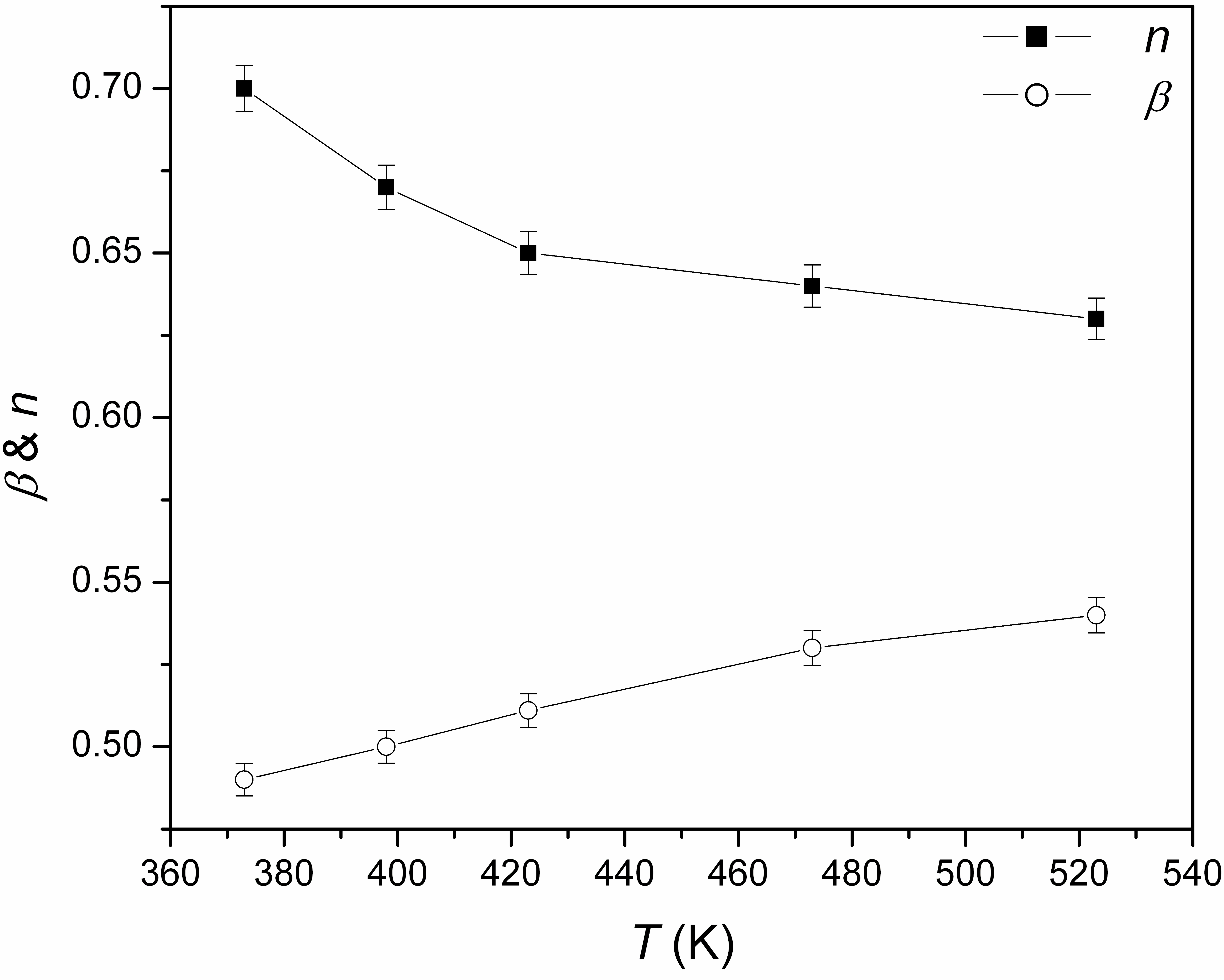

n
β
0.70
0.65
0.60
0.55
0.50
β & n
360
380
400
420
440
460
480
500
520
540
T (K)

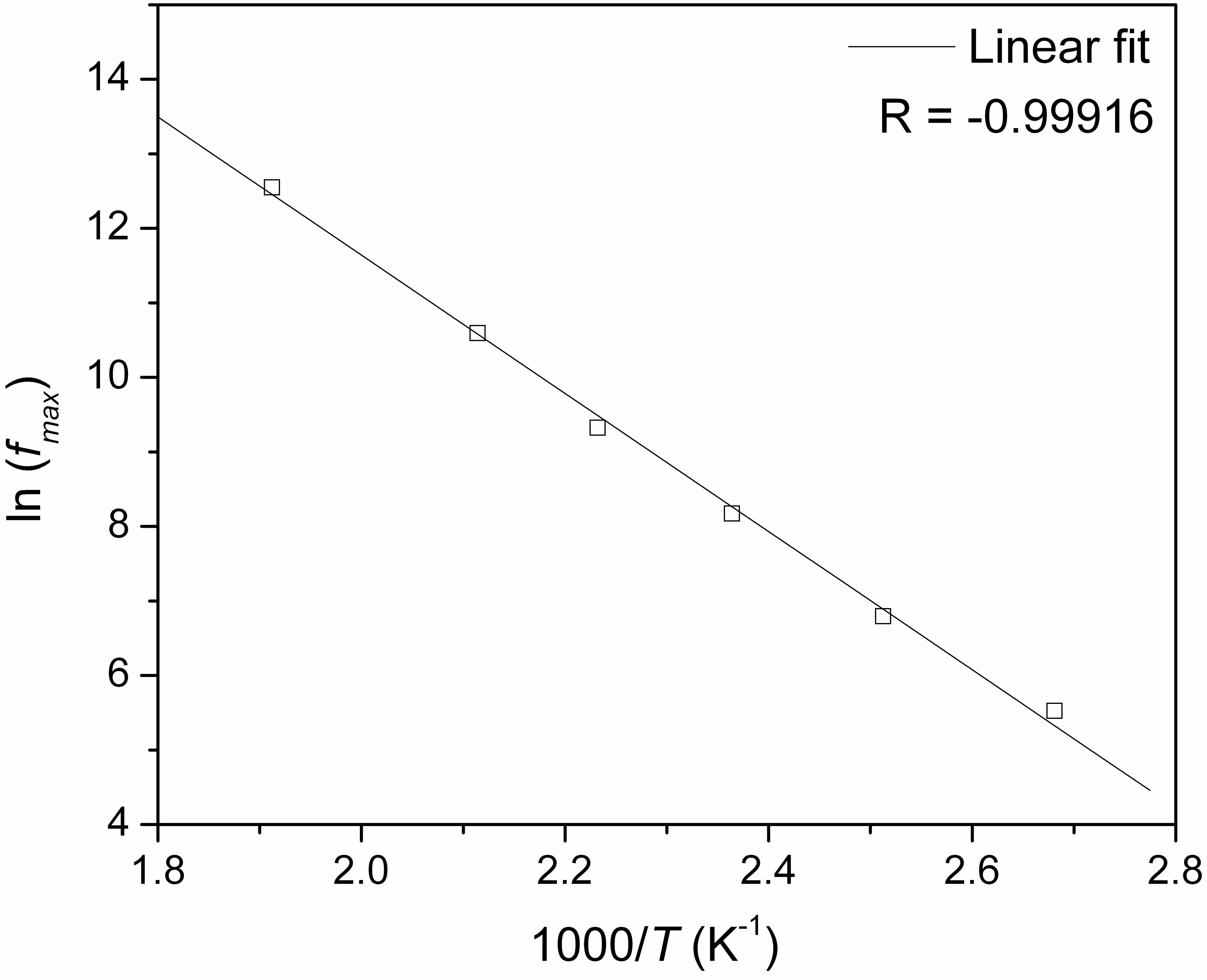

Linear fit
R = -0.99916
ln ($f_{max}$)
1000/T (K-1)
4
6
8
10
12
14
1.8
2.0
2.2
2.4
2.6
2.8

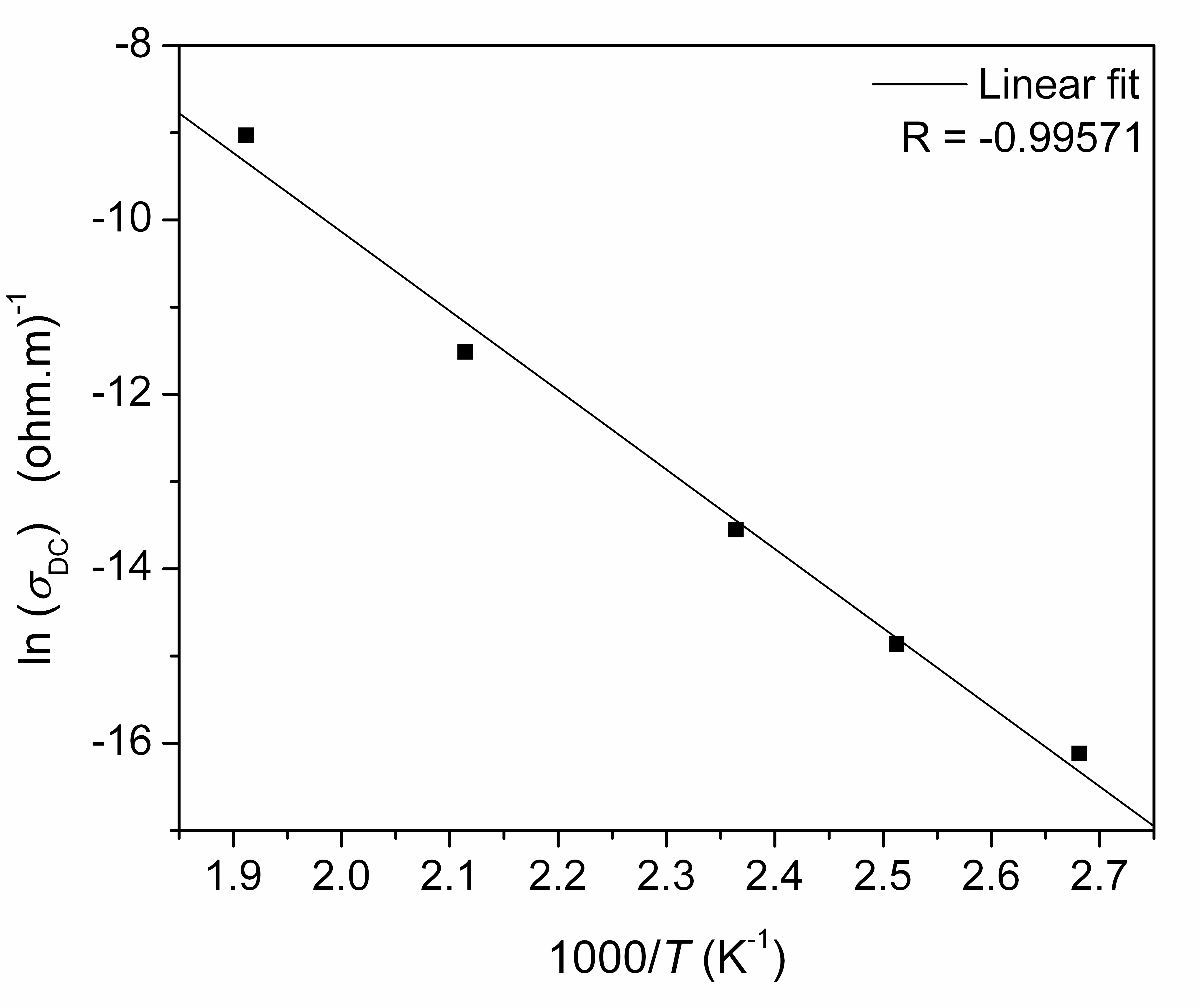

Linear fit
R = -0.99571
-8
-10
-12
-14
-16
1.9
2.0
2.1
2.2
2.3
2.4
2.5
2.6
2.7
$\ln(\sigma_{DC})$ (ohm.m)$^{-1}$
1000/$T$ (K$^{-1}$)

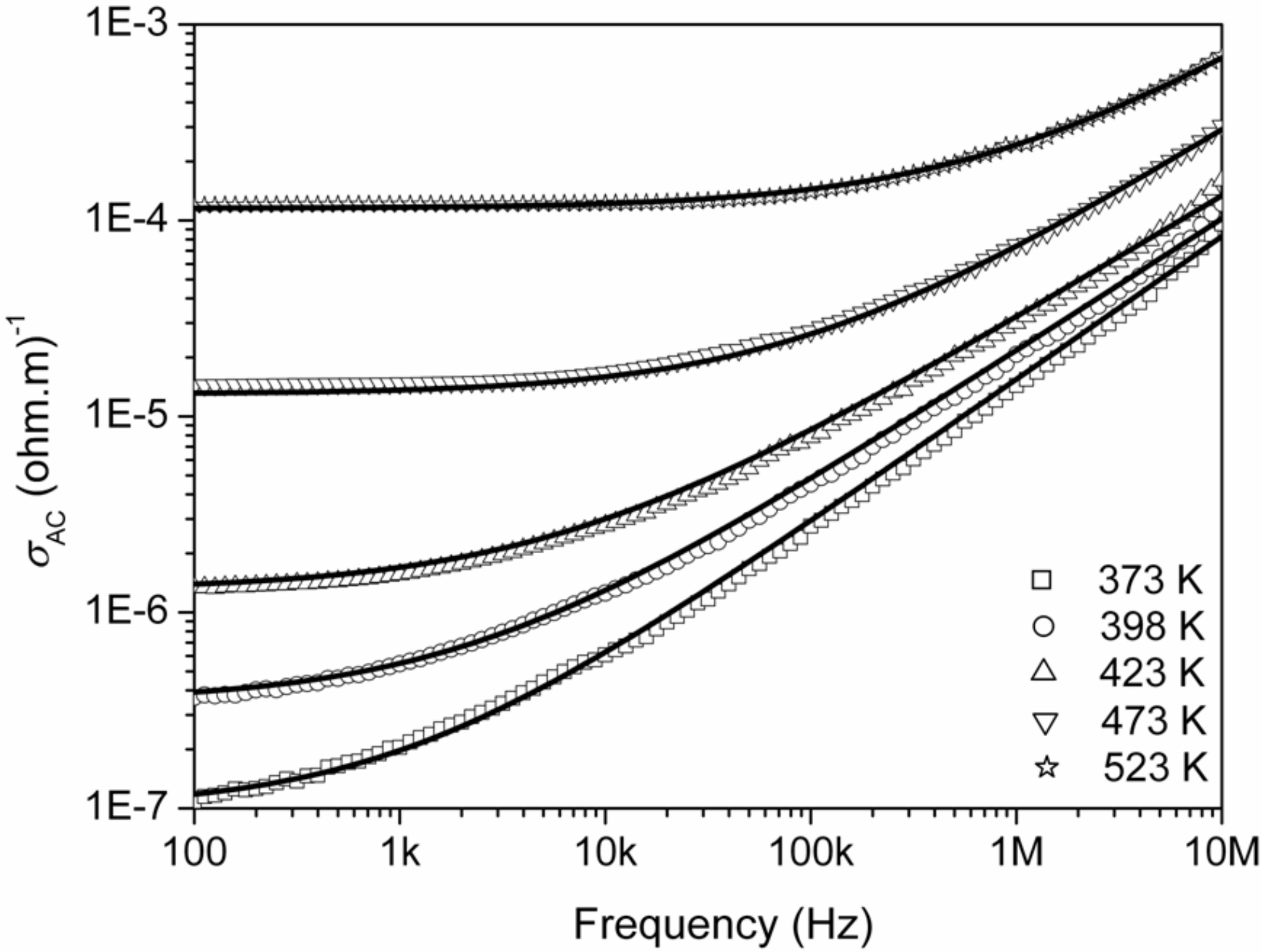

1E-3
1E-4
1E-5
1E-6
1E-7
100
1k
10k
100k
1M
10M
Frequency (Hz)
$\sigma_{AC}$ (ohm.m)$^{-1}$
373 K
398 K
423 K
473 K
523 K

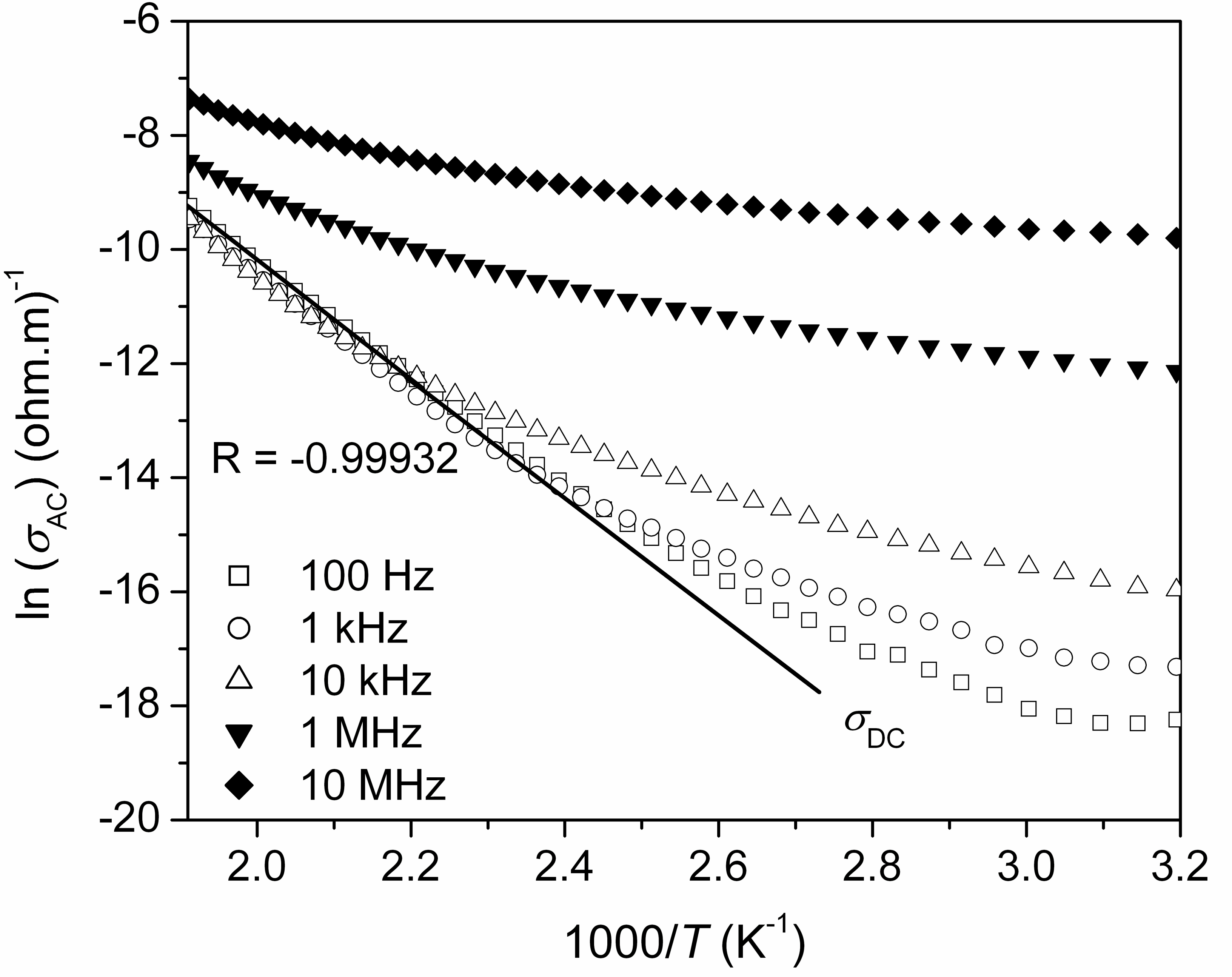

-6
-8
-10
-12
-14
-16
-18
-20
2.0
2.2
2.4
2.6
2.8
3.0
3.2
ln ($\sigma_{AC}$) (ohm.m)$^{-1}$
1000/$T$ (K$^{-1}$)
R = -0.99932
100 Hz
1 kHz
10 kHz
1 MHz
10 MHz
$\sigma_{DC}$